\documentclass[reprint,aps,
prb]{revtex4-2}

\usepackage{graphicx}
\usepackage{bm}
\usepackage{pgfplots}
\pgfplotsset{compat=1.18}
\usepackage{amsfonts}
\usepackage{amsmath}
\usepackage{amssymb}
\usepackage[utf8]{inputenc}
\usepackage[english]{babel}
\usepackage{xcolor}
\RequirePackage[T1]{fontenc}
\usepackage{appendix}
\usepackage{float}
\usepackage{braket}
\usepackage{relsize}
\usetikzlibrary{arrows}
\usepackage{enumerate}
\usepackage{newfloat}
\AtBeginDocument{\RenewCommandCopy\qty\SI}  
\usepackage{physics}
\usepackage{mathtools}
\usepackage{comment}
\usepackage{nicematrix}
\usepackage[none]{hyphenat} 
\usepackage[colorlinks=true, allcolors=blue]{hyperref}

\begin{document}
\preprint{}

\title{Gate-tunable spin splitting for Andreev spin qubits in Ge quantum well heterostructures}

\author{Avani Vivrekar}
\affiliation{Department of Physics, University of Wisconsin-Madison, Madison, Wisconsin 53706, USA}
\author{Benjamin D. Woods}
\affiliation{Department of Physics, University of Wisconsin-Madison, Madison, Wisconsin 53706, USA}
\author{Mark Friesen}
\altaffiliation[Contact author: ]{friesen@physics.wisc.edu}
\affiliation{Department of Physics, University of Wisconsin-Madison, Madison, Wisconsin 53706, USA}

\begin{abstract} 
Ge-based super-semi heterostructures are a promising platform for implementing Andreev spin qubits (ASQs).
For example, they can be isotopically purified to eliminate nuclear spins that cause decoherence. 
However, spin-orbit coupling (SOC), which is a key ingredient for ASQs, is cubic in momentum for hole-based materials like Ge, in contrast with more-conventional ASQ materials (\emph{e.g.}, InAs wires), for which the SOC is linear in momentum. 
In this work, we study the interplay between geometrical properties like the channel width, and gate-tunable properties like the chemical potential, on ASQ operations.
Intriguingly, we find that the spin splitting in a planar Ge Josephson junction does not depend on the junction width, in contrast with InAs nanowires. 
We explore the possibility of exploiting this property to suppress the harmful effects of charge traps.
However, this prospect is spoiled by the fact that charge traps also affect the chemical potential, in addition to the channel width.  
Our results provide insights into the fundamental mechanisms governing spin splitting in ASQs and offer guidance for optimizing ASQs in other materials with cubic SOC.
\end{abstract}
\maketitle

\section{Introduction}
\label{Introduction}
Super-semi hybrid qubits such as Andreev spin qubits (ASQs)~\cite{science.373.6553430,PitaVidal2023, PITAVIDAL20271} seek to combine the beneficial aspects of superconducting and semiconducting qubits:
like semiconducting qubits, they enable gate-voltage tuning of the Fermi level~\cite{RevModPhys.95.025003}, and like superconducting qubits, they allow phase tuning of the energy levels~\cite{PhysRevA.76.042319}.  
ASQs were initially proposed in the context of Josephson junctions, in the constriction of a superconductor with spin-orbit coupling~\cite{PhysRevLett.90.226806, PhysRevB.77.045311, PhysRevB.81.144519}.
More recently, ASQs were also predicted to arise in proximitized semiconductor nanowires~\cite{PhysRevB.96.125416}. 
Although ASQs are the focus of the present work, these super-semi hybrids also form the basis for other qubit implementations like topological qubits~\cite{tq1, tq2, tq3, tq8, tq4, tq9,  tq7, tq6, tq5} and gatemons \cite{larsen2015semiconductor, de2015realization, Kiyooka_2024}.

\begin{figure*}[t] 
    \centering
    \includegraphics[width=0.75\textwidth]{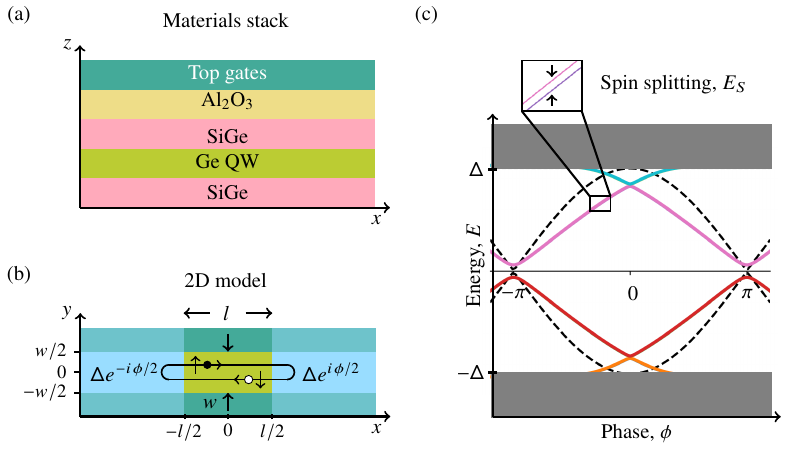}  
    \caption{
    The Ge-based Josephson junction device considered in this work and the energy dispersion of the Andreev bound states (ABSs) it can host.
    (a) A cross-sectional view of a Ge quantum-well materials stack. 
    (Layer thicknesses are not drawn to scale.)
    A thin strained Ge layer is sandwiched between much thicker layers of strain-relaxed SiGe.
    Additional layers include an Al$_2$O$_3$ oxide and nanofabricated metal top gates.
    A superconductor may be annealed into the structure from above (not shown here), inducing a proximitized superconducting gap $\Delta$ in the Ge, where the superconductor makes contact with hole carriers in the quantum well.
    (b) A top-down view of the device, showing top gates used for accumulation (green) and depletion (teal), and two superconducting leads (blue), with tunable superconducting phases, $\pm\phi/2$, which are annealed to make contact with the Ge quantum well, as described above.
    A Josephson junction of length $l$ and width $w$ forms below the accumulation gate, where the value of $w$ is tuned by the voltages applied to the depletion gates.
    The loop depicts an ABS in the junction, comprised of a particle (black circle) and an antiparticle (white circle), with opposite spins, traveling in opposite directions.
    (c) Energy vs phase relationships for ABSs formed in the Josephson junction.
    The black dashed line describes the short-junction limit, which occurs when $l$ is much smaller than the superconducting coherence length $\xi$.
    The solid, colored lines show solutions for a longer junction, with parameters given by $l=300$~nm, $\Delta=0.165$~meV, and $w=75$~nm. 
    In this calculation, we assume a hole effective mass of $m_\text{eff}=0.023m_e$, a transverse ($\hat y$) confinement energy $\hbar\omega_y=2.35$~meV, and a cubic SOC coefficient $\alpha_C = 400$~meV$\cdot$nm$^3$.
    The inset shows the SOC-induced spin splitting, away from the spin-degenerate points at $\phi=0,\pm\pi$.}
    \label{fig:fig1}
\end{figure*}

ASQs have been successfully implemented in proximitized semiconductors, most prominently in InAs nanowires~\cite{science.373.6553430, PitaVidal2023}. 
Recent work has shown that decoherence in this system is primarily of magnetic origin~\cite{hoffman2024decoherenceandreevspinqubits}, and is limited by the intrinsic spinfulness of the InAs nuclei~\cite{science.373.6553430}. 
It is therefore of interest to consider alternative spinless materials platforms.

Ge stands out as a promising potential platform for ASQs since isotopic purification can yield extremely low densities of nuclear spins \cite{Scappucci2021}.
Theoretical work has clarified the nature of proximity-induced superconductivity in Ge heterostructures~\cite{PhysRevB.108.155433, babkin2024superconductingproximityeffecttwodimensional, pino2024theorysuperconductingproximityeffect}. 
In contrast with InAs, the Fermi level of metals in contact with Ge aligns closely with the top of the valence band~\cite{dimoulas2006fermi, de2015realization}, enabling superconducting junctions with holes as the fundamental carriers \cite{Sammak2019}. 
Several experiments have demonstrated a hard induced gap with near-unity transparency that could potentially host ASQs~\cite{PhysRevResearch.3.L022005, PhysRevB.108.155433,  Kiyooka2026, Hendrickx2019}.
For example, a gate-tunable transmon was recently implemented in a proximitized Ge heterostructure~\cite{Kiyooka_2024,Sagi2024}.  
Planar Ge has also been proximitized by a platinum germanosilicide (PtSiGe) superconductor to form a quantum dot, coupled between two superconducting leads~\cite{Lakic2025}. 
Generally, Ge is expected to provide an excellent platform for hosting quantum devices~\cite{Scappucci2021, Thompson2026}.

Spin-orbit coupling (SOC) is the second ingredient needed for ASQs, together with superconductivity.
As discussed below, the dependence of SOC on momentum is cubic in planar Ge ($\sim$$k^3$), as opposed to InAs, where its dependence is linear ($\sim$$k$)~\cite{Winkler2003, Luethi2023}. 
In this paper, we explore some consequences of this distinction, and how it affects the physics of Andreev bound states (ABSs), which are discrete states in the Josephson junction that can host ASQs~\cite{Andreev1964, Kulik1970, Hays2021, PhysRevB.77.045311, PhysRevB.81.144519, PhysRevLett.90.226806}.
In experiments to date, we note that it has not been possible to resolve the spin-splitting spectra of ABSs in Ge~\cite{PRXQuantum.5.030357}, due to its weaker SOC, compared to InAs.
Thus, we also explore whether the inherent tunability of Ge devices, including parameters like the chemical potential and the electrostatically defined channel width, can be used to tune the SOC, to enhance the spin energy splitting. 

The paper is organized as follows. 
In Sec.~II, we briefly summarize the physics of ABSs, as relevant for the present discussion, and we describe our theoretical model.
In Sec.~III, we report our results, leaving technical details for the appendices. 
First, we describe the dependence of the ABS zero-field spin splitting on different system parameters, including geometrical parameters (especially the width of the Josephson junction formed in the semiconductor), and materials parameters (especially the SOC), for both the cubic and linear cases. 
Here, we also take the opportunity to discuss the origin of spin splitting and its relation to the subband physics, and we employ an effective one-subband minimal model to explain the observed power-law dependence of the spin splitting on SOC constants. 
Second, we describe the effects of a nearby trapped charge on the spin splitting.
Our simulations reveal an unexpected result, which we explain by exploring the separate effects of the chemical potential and channel confinement on the spin splitting. 
Finally, in Sec.~IV, we conclude.

\section{Model}
\label{Model}

Figure~\ref{fig:fig1} shows the model geometry considered in this work.
Figure~\ref{fig:fig1}(a) is a simplified cross-sectional view of the heterostructure, comprised of a strained Ge quantum well sandwiched between strain-relaxed, Ge-rich SiGe layers.
Other features include an insulating Al$_2$O$_3$ oxide layer, with nanofabricated top gates, whose potentials are tuned to accumulate carriers and modify the electrostatic landscape.
The quantum well is type-I, which allows for the capture of holes~\cite{Schaffler1997, PhysRevB.103.125201, Sammak2019}.
Additionally, for most metals in contact with Ge, the Fermi level is pinned very near the valence-band edge \cite{Sammak2019}, favoring the formation of a two-dimensional hole gas (2DHG) at the upper Ge/SiGe interface.
Except for the junction region, a superconductor is annealed into the heterostructure, inducing a proximitized superconducting gap of magnitude $\Delta$ in the 2DHG.

Figure~\ref{fig:fig1}(b) shows a top view of the device.
The individually tunable metal gates accumulate a channel of width $w$ and length $l$ in the plane of the 2DHG, inducing a Josephson junction between the two superconductors.
Here, the top gate primarily controls the chemical potential ($\mu$) of the channel, while the side gates deplete the channel barriers and control the channel width.
For convenience, we assume harmonic confinement along the transverse axis, $\hat y$.

For completeness, we now briefly summarize the physics of ABSs.
At an S-N junction, single-particle excitations with energies below the superconducting gap are allowed in the normal region (N), but not the superconductor (S).
In an S-N-S Josephson junction, an excited state with energy $|E| < \Delta$ is therefore bound, like a particle in a box.
However, the boundary conditions at the sides of the box follow the rules of Andreev reflection, in which a particle is reflected as an antiparticle and vice-versa.
(We use the terminology ``particles'' and ``antiparticles'' to refer to Nambu particles and Nambu holes, respectively, to avoid confusion with electrons and holes.)
The particle undergoing reflection gains an energy-dependent phase as well as a superconducting phase $\phi$.
The resulting ABS can be described as a constructive interference effect, occurring when the phase change around the loop shown in Fig.~\ref{fig:fig1}(b) is an integer multiple of $2\pi$.
This interference condition can be expressed as~\cite{Hays2021, monaghan2024rolerashbaspinorbitinduced}
\begin{equation}
    -2\arccos{\left(\frac{E}{\Delta}\right)}\pm \phi +2k_xl= 2\pi n ,
    \label{eq:interference}
\end{equation}
where $E$ is the bound-state energy, $2k_xl$ is the quantum ``dynamical'' phase gained while traversing the loop with momentum $\hbar k_x$, and $n$ is an integer.
Note that we have assumed transparent super-semi interfaces here, for simplicity.

Equation~(\ref{eq:interference}) describes an implicit dispersion relation for $E(\phi)$.
For example, if $E$ has purely kinetic origins, and if we linearize $E$ near the Fermi level, then we can substitute 
\begin{equation}
E\approx\hbar k_xv_F ,
\label{eq:vF}
\end{equation}
where $v_F$ is the Fermi velocity.
In the short-junction limit ($l\rightarrow 0$), we can easily solve this equation to obtain the cosine dispersion relation, shown with black dashed lines in Fig.~\ref{fig:fig1}(c).
For longer junctions, the solid, colored lines in the figure show energy dispersions obtained from the full 2D model described below.
Similar to a particle in a box, the energy spacings decrease as a function of (increasing) box size ($l$), drawing higher bound-state energy bands into the gap.

The quantum information of an ASQ is stored in its spin, so the spin energy splitting $E_S$ is of particular interest.
Such splittings are allowed in a superconductor because its complex phase breaks time-reversal symmetry, except at the special points where $\phi$ is an integer multiple of $\pi$~\cite{PhysRevB.77.045311, PhysRevB.81.144519, PhysRevLett.90.226806}.
SOC provides the mechanism for splitting the spin states, as illustrated in the inset of Fig.~\ref{fig:fig1}(c) (except at points where time-reversal symmetry is enforced).
In this work, we explore the possibility of using the transverse confinement $w$, which can be controlled electrostatically, to tune the SOC, and thereby $E_S$.

We now summarize our theoretical model for a Ge Josephson junction.
In our simulations, we adopt Ge materials parameters, including for cubic SOC. 
For comparison purposes, we also consider linear SOC, taking the InAs materials system as an example, as described in Sec.~\ref{InAsModel}.

\subsection{Modeling Germanium}
\label{GeModel}

To explore the physics of ABSs, we consider a simplified geometry, which includes the key features of the full 3D geometry shown in Fig.~\ref{fig:fig1}.
Specifically, we ignore the macroscopic features of the superconductor outside the Ge quantum well and limit our considerations to the proximitized portion of the device, inside the quantum well.
(This is a common approximation used in many ABS models \cite{tq6, tq10, tq3, tq11}.)
The voltages applied to the top gates are then reflected in the chemical potential $\mu$ and the channel width $w$.
As described in Appendix~\ref{AppendixD}, we begin by describing the 2DHG via a Luttinger-Kohn (LK) Hamiltonian in three spatial dimensions, including the effects of strain.
Since the split-off hole band is distant in energy ($\sim$290~meV), we adopt the 4D version of the LK Hamiltonian, which includes only the heavy- and light-hole bands. 
Now, the combination of strain and confinement splits these two bands \cite{PhysRevB.103.125201}, and the strong $\hat z$ confinement further splits the quantum-well eigenstates. 
Since these $z$-state splittings are larger than any of the energy scales in the Hamiltonian described below, we apply a Schrieffer-Wolff transformation in Appendix~\ref{AppendixD}, to project the system onto the ground-state $z$-orbital subspace of the heavy-hole band.
In the discussion below, the word ``subband'' is therefore reserved strictly for confinement along $\hat{y}$ (not $\hat{z}$).

The resulting Hamiltonian is effectively 2D, given by
\begin{equation}
    H = H_0 + H_{\text{SC}} + H_{\text{SOC}}
    \label{eq:H_tot} .
\end{equation}
Here, $H_0$ describes the spin-degenerate part of the Hamiltonian, defined as 
\begin{equation}
    H_0 = \left[\frac{\hbar^2}{2m_\text{eff}}(k_x^2+k_y^2) - \mu + \frac{1}{2}m_\text{eff}\omega_y^2 y^2 \right]\tau_z ,
    \label{eq:H_0}
\end{equation}
where $m_\text{eff}= 0.05 m_{e}$ is the effective mass obtained from the projection procedure in Eq.~(\ref{eq:meff_ge}), $k_{x,y}=-i \partial_{x,y}$ are momentum operators ($\hbar \mathbf{k}$ is the momentum), and $\tau_z$ is a Pauli matrix acting on the Nambu particle-antiparticle space spanned by $\Psi(\mathbf{r}) = \bigl(\psi_{\uparrow}(\mathbf{r}),\, \psi_{\downarrow}(\mathbf{r}),\, \psi^{\dagger}_{\uparrow}(\mathbf{r}),\, \psi^{\dagger}_{\downarrow}(\mathbf{r})\bigr)^{\mathrm{T}}$,
where $\psi_{\sigma}$ ($\psi^{\dagger}_{\sigma}$) are fermionic  field operators that annihilate (create) particles with spin 
$\sigma \in \{\uparrow,\downarrow\}$ at position $\mathbf{r}$.
We note that the first two components of $\Psi(\mathbf{r})$ correspond to particles, and the second two components correspond to antiparticles \cite{tq10}.
Throughout this work, we assume a fully separable confinement potential, in all three spatial dimensions, which we have already exploited by projecting the Hamiltonian onto the $z$-orbital ground state. 
The transverse confinement is expressed in the final term of Eq.~(\ref{eq:H_0}).
As noted above, we approximate this $y$ confinement as harmonic, through the relation $\omega_y = 2 \hbar/(m_\text{eff}w^2)$, between the orbital confinement energy $\hbar\omega_y$ and the effective channel width $w$.
We do not consider any confinement along $\hat x$, except from the inhomogeneous superconducting gap, as described by the superconducting Hamiltonian~\cite{PhysRevB.96.125416},
\begin{equation}
    H_{\text{SC}} = -\tilde{\Delta}(x)[\tau_{y}\cos{\tilde{\phi}(x)} + \tau_{x}\sin{\tilde{\phi}(x)}]\sigma_{y} ,
    \label{eq:H_sc}
\end{equation}
where $\sigma_y$ is a Pauli spin matrix.
Here, the proximitized superconducting gap is defined as
\[
\tilde{\Delta}(x) = 
\begin{cases} 
 \Delta, & \text{if } |x| \geq \frac{l}{2}, \\
 0, & \text{otherwise} ,
\end{cases}
\] 
where we set $\Delta= 0.165$~meV, as appropriate for aluminum~\cite{PhysRevB.96.125416}, and the superconducting phase is defined as 
\[
\tilde{\phi}(x) = 
\begin{cases} 
 -\frac{\phi}{2}, & \text{if } x < -\frac{l}{2}, \\
 +\frac{\phi}{2}, & \text{ if } x > \frac{l}{2}.
\end{cases}
\] 
The phase difference across the junction is tunable, via an external magnetic field, which we assume to be external to the device.
Unless stated otherwise, we set $\phi=\pi/2$ in the simulations reported here, such that the spin splitting is approximately maximized. 
Note again that the effective Hamiltonian, Eq.~(\ref{eq:H_tot}), includes just the (2D) heavy-hole band, whose total angular momentum quantum numbers are $j=3/2$ and $m_j=\pm 3/2$.
To clarify the spin notation in Eq.~(\ref{eq:H_sc}), we also note that $\sigma_y$ acts in the usual way on the spin basis states, $\ket{\uparrow}=\ket{+3/2}$ and $\ket{\downarrow}=\ket{-3/2}$, such that 
$\sigma_y\ket{\uparrow}=i\ket{\downarrow}$ and $\sigma_y\ket{\downarrow}=-i\ket{\uparrow}$.

\begin{figure*}[t] 
    \centering
    \includegraphics[width=0.7\textwidth]{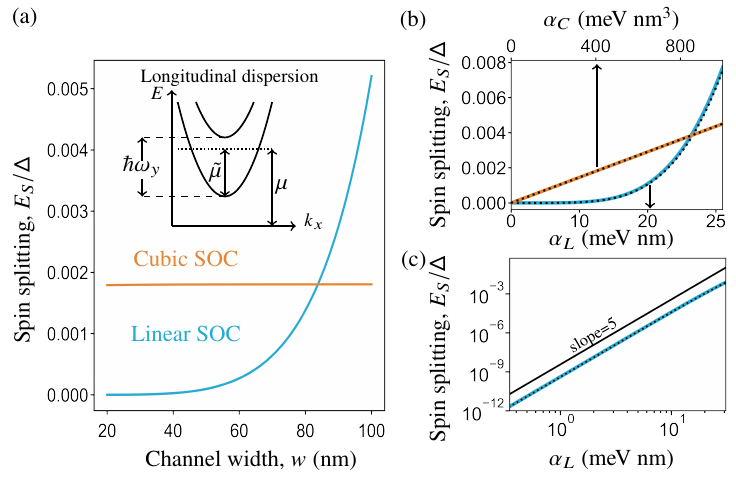} 
     \caption{Dependence of the spin splitting on channel width and spin-orbit coupling (SOC) constants. 
     (a) Dependence of the spin splitting $E_S$ (scaled by the superconducting gap $\Delta$) on the channel width $w$, for both linear and cubic SOC models.
     The inset shows the dispersion of the first two subbands in the $\hat y$ direction, with spacing $\hbar\omega_y$.
     Here, $\tilde\mu$ defines the chemical potential relative to the bottom of the lowest subband, and $\mu$ is chosen such that only a single subband is occupied: $\mu = \hbar\omega_y/2 + \tilde{\mu}$, where $\tilde{\mu}=0.8$ meV and $\hbar\omega_y=2.35$~meV.
     In the main figure, $E_S$ is computed using the same parameters as Fig.~\ref{fig:fig1}(c).  
     (b) Dependence of $E_S$ on the linear (cubic) SOC constants $\alpha_L$ ($\alpha_C$). 
     Here, the solid-blue (orange) curves show results obtained from effective 1D models, while the black dotted lines show results obtained from full 2D models, indicating nearly perfect agreement. 
     Note that the orange line is linear: $E_S\propto \alpha_C$.
     (c) Log-log plot of the same blue and dotted-black curves shown in (b); the solid-black line has slope 5, indicating that $E_S\propto \alpha_L^5$.
     Note that all solid and dotted lines in (b) and (c) represent numerical solutions.
    }
    \label{fig:fig2}
\end{figure*}

The final term in Eq.~(\ref{eq:H_tot}) describes the SOC induced by a vertical electric field $F_z$, imparted by the top gates, which couples the $z$-orbital states.
The Schrieffer-Wolff transformation described in Appendix~\ref{AppendixD} then yields the effective interaction
\begin{equation}
    H_\text{SOC} = \alpha_C[(k_x^3-3k_xk_y^2)\sigma_y\tau_z + (k_y^3-3k_x^2k_y)\sigma_x\tau_0] ,
    \label{eq:H_soc_c}
\end{equation}
where $\alpha_C$ is the effective cubic ($C$) SOC coupling constant defined in Eq.~(\ref{eq:cubic_soc_const_ge}).
This form of the SOC is necessarily cubic in momentum ($\sim$$k_{x,y}^3$), because it must couple heavy-hole states with $J_z=\pm 3/2$.
We note that, when expressed in terms of just the ``particle'' states (i.e., not the full Nambu particle-antiparticle space), Eq.~(\ref{eq:H_soc_c}) takes the familiar Rashba form, $i\alpha_C (k_{+}^3 \sigma_- - k_{-}^3\sigma_{+})$, where $k_{\pm} = k_x \pm ik_y$ and $\sigma_{\pm} = (\sigma_x \pm i\sigma_y)/2$.
In this work, we adopt typical values of electric field and confinement energies, as explained in Appendix~\ref{AppendixD}, leading to an estimate of $\alpha_C=400$~meV$\cdot$nm$^3$. 
Equations~\eqref{eq:H_tot}-\eqref{eq:H_soc_c} describe a set of coupled partial differential equations, which we discretize and solve numerically on a 2D grid. 

To gain further insight into the dependence of the spin splitting on channel width, we also derive an effective 1D model by projecting the 2D Hamiltonian onto the $\hat y$ harmonic-oscillator eigenstates defined in Eq.~\eqref{eq:H_0}.
The full Hamiltonian is eventually reduced to a 1D Hamiltonian, as described in Appendix~\ref{AppendixA},
expressed as
\begin{equation}
    H_\text{1D} = \left(\frac{\hbar^2 k_x^2}{2m^*_\text{eff}} - \mu^*\right)\sigma_0 \tau_z 
    +H_\text{SC}
    +(\alpha_L^* k_x + \alpha_C^* k_x^3)\sigma_z\tau_0 ,
    \label{eq:H1D}
\end{equation}
where $m^*_\text{eff}$, $\mu^*$, $\alpha_L^*$, and $\alpha_C^*$ are renormalized Hamiltonian parameters.
Note that the superconducting term $H_\text{SC}$ is unaffected by this 1D projection.
For the case of cubic SOC, the Hamiltonian is simply truncated at the lowest $y$ subband, yielding an equally simple set of Hamiltonian parameters, as will be explained in Sec.~\ref{ResultsA}. 
Equation~\eqref{eq:H1D} is then solved by discretizing on a 1D grid along $\hat x$.
Below, we compare results from these 1D and 2D models, obtaining essentially perfect agreement, with much less effort required in the 1D case.

The models described above were developed to provide an accurate representation of Josephson junctions in the Ge material system.
However, a main goal of this work is to provide a comparison with the more-established InAs materials system.
To aid this comparison, we now replace the Ge effective mass in Eqs.~\eqref{eq:H_tot}-\eqref{eq:H1D} with the InAs effective mass, given by $m_\text{eff}=0.023m_e$.
In contrast to InAs however, we retain the cubic form for the SOC in Ge.
For example, the dispersion relation plotted in Fig.~\ref{fig:fig1}(c) was computed in this way, using the InAs effective mass.

\subsection{Modeling InAs}
\label{InAsModel}
As noted, for comparison purposes, we also study Josephson junctions in planar InAs quantum wells \cite{PhysRevApplied.19.054026}.
In this case, the carriers are in the conduction band and the SOC is linear in momentum, since it couples states with $J_z=\pm 1/2$~\cite{Winkler2003}.
Equation~(\ref{eq:H_soc_c}) is then replaced by
\begin{equation}
    H_\text{SOC} = \alpha_L (-k_x\sigma_y\tau_z + k_y\sigma_x\tau_0) ,
    \label{eq:H_lsoc}
\end{equation}
where $\alpha_L$ is the linear ($L$) SOC coefficient, which takes a typical value of $\alpha_L=20$~meV$\cdot$nm for InAs~\cite{PhysRevB.96.125416}.
Except for this difference in $H_\text{SOC}$, the theoretical descriptions in InAs are the same as Ge.
However, to capture the leading-order effects of the linear SOC, we note that projection onto a 1D theoretical model requires that we include couplings to the excited $y$ subbands, as described in Appendix~\ref{AppendixA}.
In this case, we then perform a Schrieffer-Wolff transformation to block diagonalize the system with regard to the $y$ subbands, and finally reduce the problem to the lowest subband.
The resulting effective 1D Hamiltonian has the same form as given in Eq.~(\ref{eq:H1D}), although the renormalized Hamiltonian parameters are modified, as will be discussed in Sec.~\ref{ResultsA}.

\begin{figure*}[t]  
    \centering
    \includegraphics[width=0.6\textwidth]{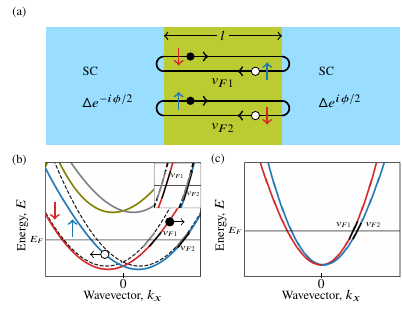}  
    \caption{
    Origins of spin splitting for systems with linear vs cubic SOC.
    (a) A cartoon depiction of ABSs, where the loops represent solutions to the constructive interference condition, Eq.~(\ref{eq:interference}).
    [See also Fig.~\ref{fig:fig1}(b).]
    These solutions describe a discrete set of $k_x$ values (or equivalently, $E$ values) that depend on the Fermi velocity $v_F$ through Eq.~(\ref{eq:vF}).
    If the two bound states have different $v_F$, this gives a small energy splitting, known as the ``spin splitting,'' $E_S$.
    (b) Energy dispersions for spin states with linear SOC.
    The lower bands (colored or dashed curves) correspond to the ground $y$ subband, while the upper bands (gray curves) correspond to excited states.
    A particle ($E>E_F$, black circle) and an antiparticle ($E<E_F$, white circle) combine to form an ABS when $k_x$ satisfies Eq.~(\ref{eq:interference}).
    When the spin-orbit couplings between subbands are turned off, the resulting dispersions (dashed curves) describe identical, shifted  parabolas, with the same $v_{F1}=v_{F2}$, because their slopes at $E=E_F$ are equal.
    When subband couplings are turned on (red and blue curves), the dispersions are distorted, giving $v_{F1}\neq v_{F2}$ and a spin splitting.
    (c) Energy dispersions for spin states with cubic SOC.
    In this case, the cubic coupling gives $v_{F1}\neq v_{F2}$ and a spin splitting, even without couplings to excited subbands.
    }
    \label{fig:fig3}
\end{figure*}

\section{Results}
\subsection{Dependence of the Spin Splitting on System Parameters}
\label{ResultsA}
The main goal of this work is to explore the dependence of the spin splitting ($E_S$) of an ABS on the channel width ($w$).
Since (i) the spin splitting arises from SOC, (ii) SOC is affected by the lateral confinement strength \cite{PhysRevB.104.115425}, and (iii) cubic SOC is nonlinear in $k$, we might expect to observe a strong variation of $E_S(w)$ in Ge, as compared to a linear SOC material like InAs.
We study this possibility in Fig.~\ref{fig:fig2}(a) using the 1D theoretical model described above.
Here, we vary $w$ while keeping all other parameters fixed, including the chemical potential, $\tilde \mu$, which is defined relative to the bottom of the lowest $y$ subband, as illustrated in the inset of Fig.~\ref{fig:fig2}(a).
However, the observed results are opposite of the expected behavior.
Indeed, we find that Ge's spin splitting is essentially independent of channel width, while InAs's spin splitting grows dramatically.
Additionally, for the materials parameters considered here, InAs's spin splitting can be much larger than Ge's, which may be advantageous for gate operations in InAs.
Below, we use the 1D model to show that these differences can be explained by the nature of the couplings between ground and excited $y$ subbands, caused by cubic vs linear SOC. 
We also show that the flat behavior of $E_S$ for Ge can be attributed to the fact that we hold $\tilde \mu$ constant while sweeping $w$.

Another important contributor to the physics of spin splitting is the dependence of $E_S$ on the SOC coefficients, $\alpha_C$ or $\alpha_L$.
These parameters are materials-dependent and difficult to tune (except perhaps via strain engineering).
In Fig.~\ref{fig:fig2}(b), we plot the spin splitting for the Ge and InAs systems (orange and blue curves, respectively), revealing vastly different behaviors, including a linear dependence for Ge (cubic SOC) and a fifth-power dependence for InAs (linear SOC) [see Fig.~\ref{fig:fig2}(c)].
We now show that these behaviors can also be explained by the different linear and cubic SOC between the $y$ subbands.

We first explain the different spin-splitting behaviors observed in Fig.~\ref{fig:fig2}(a), for linear vs cubic SOC materials.
In Sec.~\ref{Model}, we noted that (i) the bound-state energy depends on the Fermi velocity $v_F$, and (ii) the slope of the energy dispersion at the Fermi level, $\partial E/\partial k_x|_F$, is proportional to $v_F$.
To compute the spin splitting, we further note that the spins of the bound states are actually pseudo-spins, composed of both spin-up and spin-down components, as illustrated in Fig.~\ref{fig:fig3}(a), where the two eigenstates are shown with different Fermi velocities.
In the absence of SOC, these (pseudo)-spin-state energy dispersions are fully degenerate, as illustrated in the inset of Fig.~\ref{fig:fig2}(a).
As shown in Figs.~\ref{fig:fig3}(b) and \ref{fig:fig3}(c), and discussed below, the application of SOC then causes the dispersions to intersect the Fermi level at slightly different $k_x$ values, with slightly different Fermi velocities.
We then see that the interference condition expressed in Eqs.~(\ref{eq:interference}) and (\ref{eq:vF}) yields slightly different spin-state energies, with a corresponding spin splitting~\cite{monaghan2024rolerashbaspinorbitinduced, Hays2021, hoffman2025resolvingandreevspinqubits, PhysRevB.96.125416}.

\begin{figure*}[t]  
    \centering
    \includegraphics[width=0.8\textwidth]{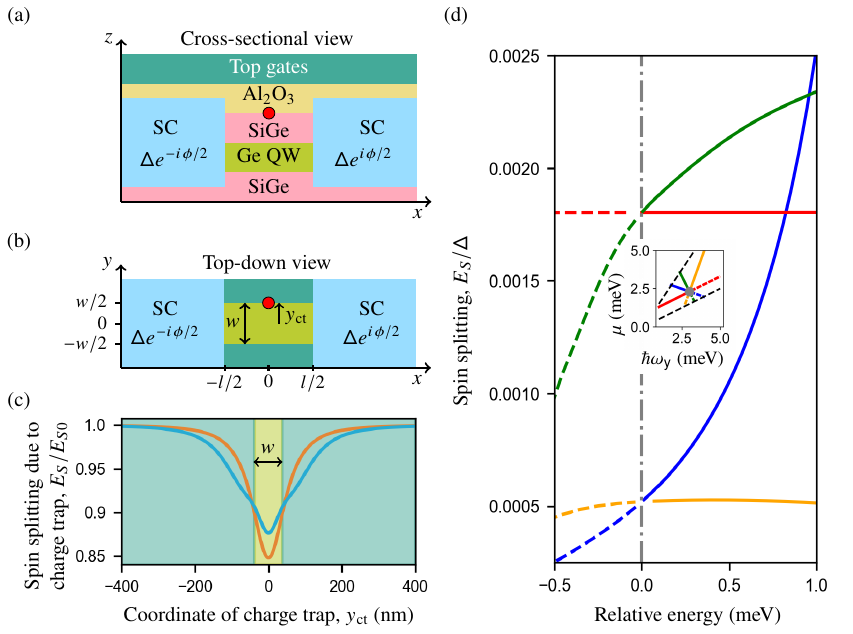}
    \caption{Effect of a charge trap on the spin splitting $E_S$. 
    (a) Defects are commonly found at the semiconductor-oxide interface, as illustrated by the red circle in a cross-sectional view of the device.
    (b) We consider a model where the defect is positioned halfway between the two superconducting leads ($x=0$), at varying locations $y=y_\text{ct}$, as indicated in this top-down view.
    (c) Effect of the charge trap on the scaled spin splitting, $E_S/E_{S0}$ ($E_{S0}$ is the spin splitting with no charge trap), as a function of $y_\text{ct}$, for both linear (blue) and cubic (orange) SOC. 
    Since the charge trap modifies the effective width of the channel, we could expect from Fig.~\ref{fig:fig2}(a) that it has no effect on the spin splitting for the case of cubic SOC.
    However, we observe here similar behaviors for both linear and cubic SOC, because the charge trap affects the chemical potential $\mu$ in addition to the channel width $w$.
    (d) To see this, we plot the scaled spin splitting $E_S/\Delta$ as a function of the relative energy, defined as $[(\mu-\mu_0)^2+(\hbar\omega_y-\hbar\omega_0)^2]^{1/2}$, along several different paths through the ($\mu,\hbar\omega_y$) parameter space, where $\omega_y = 2 \hbar/(m_\text{eff}w^2)$.
    The inset defines these paths and their point of intersection, ($\mu_0,\hbar\omega_0$) (gray circle).
    Here, the dashed black lines represent the boundaries of the region where a single $y$ subband is filled, as consistent with the inset of Fig.~\ref{fig:fig2}(a).
    For linear SOC, the yellow (blue) lines are chosen as paths with weak (strong) $E_S$ variations, as seen in the main figure.
    For cubic SOC, the red (green) lines are chosen as paths with weak (strong) $E_S$ variations.
    In particular, the red curve corresponds to the case with $\tilde\mu=\text{constant}$, as considered in Fig.~\ref{fig:fig2}(a).
    Here, the gray dot-dashed line corresponds to zero relative energy (gray circle in the inset).
    The dashed vs solid portions of the colored (non-gray) curves clarify the direction of the paths in parameter space, by comparing to the inset. 
    }
    \label{fig:fig4}
\end{figure*}

The effect of SOC on the energy dispersion and the spin splitting is illustrated in Fig.~\ref{fig:fig3}(b) for the linear case and in Fig.~\ref{fig:fig3}(c) for the cubic case.
In the linear case, the most obvious effect is the shifting of the centers of the parabolas to the left and right, while in the cubic case, the parabolas remain centered at $k_x=0$.
To explore how these behaviors affect spin splitting, we first turn off the couplings between the ground and excited $y$ subbands in the 1D effective model of Sec.~\ref{Model}.
For the case of linear SOC, we obtain the dashed lines shown in Fig.~\ref{fig:fig3}(b).
In Appendix~\ref{AppendixB}, we confirm that these two dispersions are parallel at the Fermi level (labeled $E_F$ in the figure), indicating the absence of spin splitting.
Now turning on the couplings to the excited subbands (solid red and blue curves), we observe stronger interactions between subbands that are more-closely spaced, yielding different Fermi velocities for the two spin bands, and a corresponding energy splitting.
For larger channel widths $w$, the level spacing decreases, yielding even larger spin splittings, as consistent with Fig.~\ref{fig:fig2}(a).
Now turning to cubic SOC, Fig.~\ref{fig:fig3}(c), we note that the shape of the dispersion is naturally distorted from parabolic, even without coupling to excited $y$ subbands, which represents the leading-order spin-splitting mechanism.
Our full simulations confirm that this cubic distortion is the dominant contribution to the spin splitting and that couplings to excited subbands have a negligible effect.
Thus, there is no dependence of the level spacing (or $w$) on the spin splitting in this case, as consistent with Fig.~\ref{fig:fig2}(a).
In summary, we have explained the distinct linear and cubic SOC behaviors observed in Fig.~\ref{fig:fig2}(a), and we have demonstrated that spin splitting in linear SOC materials is caused by inter-subband couplings, while for cubic SOC materials, it is an intra-subband effect.

We now explain the dependence of the spin splitting on the SOC constants, as shown in Figs.~\ref{fig:fig2}(b) and \ref{fig:fig2}(c).
In Appendix~\ref{AppendixB}, we present general arguments confirming that only cubic-in-$k_x$ terms in the effective 1D SOC can cause spin splitting, and that this splitting depends linearly on the cubic SOC coefficient $\alpha_C$.
Additionally, in Appendix~\ref{AppendixA} we note that, for cubic SOC, the prefactor of the cubic term is unaffected when we project the 2D Hamiltonian onto the 1D Hamiltonian, so that $\alpha_C^*=\alpha_C$.
($m_\text{eff}^*=m_\text{eff}$ is also unaffected by the 1D projection; we refer the reader to Appendix~\ref{AppendixA} for details on the other model parameters.)
This explains the linearity of the orange curve in Fig.~\ref{fig:fig2}(b), where the solid-orange line describes the solution obtained from the 1D model while the corresponding black dotted line shows perfect agreement with the full 2D model.
Given these results, how can we understand the observed strong dependence of $E_S$ on the linear SOC coefficient $\alpha_L$?
The answer lies in the renormalization of the linear SOC when performing the 1D projection, as discussed in Appendix~\ref{AppendixA}.
Crucially, we find that (i) an effective cubic SOC term emerges from this projection, due to the $y$ subband interactions, and (ii) its coefficient $\alpha^*_C$ is proportional to $\alpha_L^5$ [see Eq.~(\ref{eq:fifthpower})], as consistent with Figs.~\ref{fig:fig2}(b) and \ref{fig:fig2}(c). 
For the case of linear SOC, all the terms of the effective 1D Hamiltonian are renormalized, as described in Appendix~\ref{AppendixA}.

\subsection{Possible Immunity to Charge Noise}
\label{ResultsC}
As noted above, the independence of spin splitting on channel width for cubic SOC is easy to understand in Fig.~\ref{fig:fig2}(a): it reflects the fact that the spin splitting is dominated by the intrinsic energy dispersion rather than interactions between the $y$ subbands.
Nonetheless, the behavior is remarkable in comparison with linear SOC, suggesting that the spin splitting in such a system could be immune to disturbances like charge noise that affect the local electrostatic potential and the channel width.
As we show below, an important related issue is how the Fermi energy is defined in our analysis.
For the simulations reported in Fig.~\ref{fig:fig2}(a), we chose to hold fixed the chemical potential \emph{difference}, $\tilde\mu$, defined relative to the bottom of the subband.
However, in realistic devices where the potential energy fluctuates (e.g., due to charge disorder), the bottom of the $y$ subband varies locally, in contrast to the Fermi level, which is globally uniform, resulting in local variations in $\tilde\mu$.
Such variations affect the Fermi velocity $v_F$ and therefore the spin splitting, as anticipated in Appendix~\ref{AppendixB} [Eq.~(\ref{eq:spin_splitting_deltav_rel})].
It is therefore important to explore the competition between $\tilde\mu$ variations and $w$ variations (or equivalently, $\omega_y$ variations), to determine whether the flat behavior observed in Fig.~\ref{fig:fig2}(a) is truly significant.

To study this question, we consider the charge trap model shown in Figs.~\ref{fig:fig4}(a) and \ref{fig:fig4}(b). 
Here, the charge trap (indicated by the red circle) is located in the oxide region of the device -- most commonly at the semiconductor-oxide interface~\cite{ct_sim}.
Analogous to a voltage applied to a gate, this charge defect can alter the accumulation of charge carriers in the quantum well below it, effectively changing the shape and the width of the Josephson junction.
For our simulations, we assume the charge is situated halfway between the superconducting leads, at $x=0$, and we study its effect as a function of position along $y$, defined as $y_\text{ct}$.
(Here, ``ct'' stands for charge trap.)
The electrostatic potential associated with the trapped charge is taken to be that of a point charge, with appropriate screening effects from the top gate and the heterogeneous dielectric materials.
(See Appendix~\ref{AppendixC} for details.)
The simulations are performed as described in previous sections using the fully 2D model described in Sec.~\ref{Model} for both cubic and linear SOC.
Results are shown in Fig.~\ref{fig:fig4}(c), revealing similar behaviors for the two systems.
Here, to compare apples to apples, we have considered only \emph{repulsive} trapped charges, corresponding to positive charge for Ge holes and negative charge for InAs electrons.
For the cubic SOC system, in contrast with Fig.~\ref{fig:fig2}(a), we now observe strong variations of the spin splitting, which we attribute to fluctuations of the relative chemical potential $\tilde\mu$.

To further explore the dependence of spin splitting on $w$, while also varying $\mu$, we return to the simple model of Sec.~\ref{Model}, containing no charge traps.
The inset of Fig.~\ref{fig:fig4}(d) shows a map of the parameter space we explore, where the confinement energy $\hbar\omega_y$ is related to the channel width, as explained in Sec.~\ref{Model}; for reference, $\hbar\omega_y$ and $\mu$ are both illustrated in the inset of Fig.~\ref{fig:fig2}(a).
The colored lines in the inset of Fig.~\ref{fig:fig4}(d) describe ``paths'' through the $\mu$-$\hbar\omega_y$ parameter space, while the dashed black lines demarcate regions with different subband occupations, corresponding to two, one, or zero occupied subbands (top to bottom).
[In the present work, we consider only one occupied subband, as consistent with Fig.~\ref{fig:fig2}(a).]
The paths have been chosen to exemplify contrasting types of behavior, including (i) nearly constant spin splittings for linear (yellow) or cubic (red) SOC, and (ii) strongly varying splittings for linear (blue) or cubic (green) SOC.
For convenience, the paths in the inset of Fig.~\ref{fig:fig4}(d) are chosen to be linear, and to pass through a single fixed point (gray circle), labeled $(\mu,\hbar\omega_y)=(\mu_0,\hbar\omega_0)$.  
The spin splittings along these paths are plotted in the main figure using the same color scheme and the same solid vs dashed lines.
In particular, the red curve in Fig.~\ref{fig:fig4}(d) describes the same results as the orange curve in Fig.~\ref{fig:fig2}(a), where the fixed-$\tilde\mu$ constraint is reflected in the fact that the red and dashed black paths are parallel.
Note that the ``relative energy'' in this plot is defined as the distance from the gray circle, 
$[(\mu-\mu_0)^2+(\hbar\omega_y-\hbar\omega_0)^2]^{1/2}$.

In Fig.~\ref{fig:fig4}(d) (main figure), we plot the spin splitting along the paths shown in the inset, finding that this splitting can depend strongly on both $w$ and $\mu$.
The trends observed here also support the following interpretation of behavior observed in Fig.~\ref{fig:fig4}(c).
As consistent with Figs.~\ref{fig:fig3}(b) and \ref{fig:fig3}(c), we expect that larger values of $\mu$ yield larger differences in the Fermi velocities of the two spin states, and larger spin splittings.
The results in Fig.~\ref{fig:fig4}(c) can therefore be explained by noting that repulsive charges cause a local increase of the potential energy near the defect, effectively lowering the chemical potential at this location, and the spin splitting.
We can test this interpretation by reversing the sign of the charge traps in our simulations to give attractive charges.
The results of this test (not plotted here) show spin splittings with the opposite behavior, corresponding to hills rather than valleys.
From this analysis, we conclude that the strong dependence of the spin splitting on the chemical potential largely overrides any potential immunity to charge noise inferred from Fig.~\ref{fig:fig2}(a).

\section{Summary and Conclusion}
\label{Conclusion}
In summary, we have investigated the possibility of electrically tuning the spin energy splittings of ABSs in Josephson junctions, focusing on the materials system of planar SiGe/Ge/SiGe quantum wells with proximitized superconducting leads.
The natural carriers in this system are heavy holes, with a corresponding cubic-in-$k$ spin-orbit coupling.
For comparison, we also consider a linear-in-$k$ SOC, taking electrons in the conduction band of InAs as an example.
We find that the main effect of tuning the junction width \textit{w} is to modify the $y$ subband spacings. 
For linear SOC, the presence of nearby subbands distorts the energy dispersion, which in turn causes an energy splitting of the spin states at the Fermi level; in this case, larger $w$ values lead to larger spin splittings.
In contrast, for cubic SOC, the energy dispersions of the two spin bands are nondegenerate, even without any subband couplings, so their spin splittings are unaffected by $w$.
Subband couplings also explain the observed dependence of the spin splitting on the spin-orbit coupling constants, for both linear and cubic SOC, with a much stronger dependence observed in the linear case.

The fact that the spin splitting is unaffected by subband couplings in cubic SOC materials suggests that these systems may be immune to charge fluctuations and charge traps, which can modify the channel width $w$.
However, exploring this hypothesis in a realistic model geometry reveals similar spin-splitting resilience (or nonresilience) for both linear and cubic SOC. 
This unexpected behavior occurs because charge traps simultaneously influence the relative chemical potential $\tilde\mu$, in addition to $w$. 
Furthermore, mapping out the dependence of spin splitting on the chemical potential $\mu$ and subband spacing $\hbar\omega_y$ reveals distinct ``paths'' within this parameter space, where spin splitting variations range from negligible to substantial. 
This provides a practical framework for experimental tuning. 
For instance, precise electrical control over the spin splitting is crucial when performing spin-qubit gate operations via electric dipole spin resonance (EDSR)~\cite{coppini2026strainengineeringandreevspin}. Consequently, our findings clarify the robustness of tunable spin splitting in Ge-based Andreev spin qubits, offering guidance for optimizing these architectures in experimental settings.

To conclude, we emphasize that Ge quantum wells are an important new platform for several types of spin qubits.
For Andreev spin qubits, Ge provides certain advantages over other materials, such as a natural immunity to hyperfine noise and strong tunability of its electronic properties.
However, other challenges remain, such as smaller spin splittings, compared to more-conventional ASQ materials.
Possible strategies for enhancing the spin splitting include strain engineering to generate a linear-in-$k$ SOC component~\cite{PhysRevLett.131.097002, coppini2026strainengineeringandreevspin} or heterostructure engineering to enhance the magnitude of the SOC \cite{DelVecchio2026}.
Modified geometries, such as Ge nanowires, also offer the possibility of enhanced SOC~\cite{Scappucci2021,PhysRevB.84.195314, PhysRevB.97.235422}. 
Another interesting approach would be to adopt granular Al as the proximitizing superconducting material, due to its better resilience to high magnetic fields \cite{Fabris2026}. 

\section*{Acknowledgments}

We acknowledge many helpful discussions with B.\ Harpt and M.A.\ Eriksson.
This research was sponsored in part by the National Science Foundation QLCI-HQAN (Award No.\ 2016136). 
Any opinions, findings, and conclusions or recommendations expressed in this material are those of the author(s) and do not necessarily reflect the views of the National Science Foundation.
B.D.W.\ also acknowledges a fellowship from the Army Research Office under Award No.\  W911NF-22-1-0090.
The views, conclusions, and recommendations contained in this document are those of the authors and are not necessarily endorsed by nor should they be interpreted as representing the official policies, either expressed or implied, of the Army Research Office or the U.S. Government. The U.S. Government is authorized to reproduce and distribute reprints for U.S. Government purposes notwithstanding any copyright notation herein.
Some of this work
was performed using the computing resources and assistance
of the UW-Madison Center For High Throughput Computing
(CHTC) in the Department of Computer Sciences. The
CHTC is supported by UW-Madison, the Advanced Computing
Initiative, the Wisconsin Alumni Research Foundation, the
Wisconsin Institutes for Discovery, and the National Science
Foundation, and is an active member of the OSG Consortium,
which is supported by the National Science Foundation and
the U.S. Department of Energy’s Office of Science.

\section*{Data availability}
All data in this work are generated through simulations and the corresponding codes are publicly available. (See code availability statement below.)

\section*{Code availability}
The source code for reproducing the simulations and figures  is available on Zenodo \cite{zenodo}.

\appendix

\section{Effective heavy hole model for a Ge quantum well} 
\label{AppendixD}
In this Appendix, we derive an effective 2D Hamiltonian for Ge heavy holes, including SOC generated by an electric field in the $\hat z$ direction.

As noted in the main text, the split-off valence band of Ge is well separated from the heavy and light-hole bands by an energy of $\sim$290~meV.
We therefore limit our analysis to the 4D basis states $|j, m_j\rangle = \left\{ \left| \frac{3}{2}, \frac{3}{2} \right\rangle, 
\left| \frac{3}{2}, -\frac{3}{2} \right\rangle, 
\left| \frac{3}{2}, \frac{1}{2} \right\rangle, 
\left| \frac{3}{2}, -\frac{1}{2} \right\rangle \right\}$, where $m_j=\pm 3/2$ corresponds to higher-lying heavy holes and $m_j=\pm 1/2$  corresponds to lower-lying light holes.
In this basis, the Luttinger-Kohn Hamiltonian is given by~\cite{Winkler2003,Wang2021,Wang2024}
\begin{equation}
H_{\text{LK}} =
    \begin{pmatrix}
    P + Q  & 0 & S & R \\
    0 & P + Q  & R^\dagger & -S^\dagger \\
    S^\dagger & R & P - Q  & 0 \\
    R^\dagger & -S & 0 & P - Q 
    \end{pmatrix} , \label{eq:HLK}
\end{equation}
where
\begin{equation}
\begin{aligned}
P &= P_k + P_\varepsilon, \quad & Q &= Q_k + Q_\varepsilon, \\
R &= R_k + R_\varepsilon, \quad & S &= S_k + S_\varepsilon .
\end{aligned}
\end{equation}
Here, the $k$ subscript indicates kinetic energy matrix elements, while the $\varepsilon$ subscript indicates Bir-Pikus matrix elements, which describe the strain contributions. 
These terms have the following interpretations: $P_k+Q_k$ is the kinetic energy of the heavy holes, $P_k-Q_k$ is the kinetic energy of the light holes, $S_k$ is the light-hole-heavy-hole coupling for aligned spins, and $R_k$ is the light-hole-heavy-hole coupling for anti-aligned spins.  
The individual terms are given by
\begin{gather}
    P_k = \frac{\hbar^2}{2m_e} \gamma_1 (k_x^2 + k_y^2 + k_z^2), \\
    Q_k = \frac{\hbar^2}{2m_e} \gamma_2 (k_x^2 + k_y^2 -2 k_z^2), \\
    S_k = -\frac{\hbar^2}{2m_e} 2\sqrt{3}\gamma_3 k_-k_z, \\
    R_k = -\frac{\hbar^2}{2m_e} \sqrt{3} (\bar{\gamma} k_{-}^2 + \delta k_{+}^2) ,
\end{gather}
where $m_e$ is the bare mass of an electron. 
For Ge, the Luttinger parameters are given by $\gamma_1$ = 13.38, $\gamma_2$ = 4.24, and $\gamma_3$ = 5.69~\cite{Wang2024, Winkler2003}. 
To simplify the notation, we further define $\bar{\gamma} = (\gamma_2 + \gamma_3)/2$ and $\delta= (\gamma_2-\gamma_3)/2$; although here, since $\delta\ll\bar{\gamma}$, we ignore the contributions from $\delta$, which is known as the axial approximation \cite{Winkler2003}.
We also limit our analysis to the case of uniaxial strain, giving a single Bir-Pikus contribution of $Q_\varepsilon = -\varepsilon$, which results in a strain-induced energy splitting of $2\varepsilon$ between the heavy holes and light holes.

In addition to the 4D basis of spin states, we also consider a set of quantum-well $z$-orbital states, with labels $p=1,2,\dots$.
These are computed by assuming an infinite square well of width $L_z$, resulting in the extended basis set $\{|j,m_j,p\rangle\}$.
As described in the main text, we project this larger system onto the ground $z$-orbital subspace, since the $z$-orbital excitation energies are larger than any of the energy scales in Eq.~(\ref{eq:H_tot}). 
(For discussion of a related problem where multiple $z$ orbitals are retained in the analysis, along with a magnetic field, see \cite{hoffman2025resolvingandreevspinqubits}.)
Next, we introduce a vertical electric field $F_z$ to lift the inversion symmetry and generate SOC, by adding $-eF_zz$ to the diagonal terms in Eq.~(\ref{eq:HLK}), where $e=|e|$ is the charge of the hole carriers.
Finally, we project this full system onto the desired 2D basis set $\left\{ \left| \frac{3}{2}, \frac{3}{2},1 \right\rangle, 
\left| \frac{3}{2}, -\frac{3}{2}, 1 \right\rangle\right\}$, by applying a Schrieffer-Wolff transformation \cite{Winkler2003}, obtaining
\begin{equation}
    H_{\text{eff}} \approx  \frac{\hbar^2}{2m_{\text{eff}}} (k_x^2+k_y^2)\sigma_0 
    + i\alpha_C (k_{+}^3 \sigma_- - k_{-}^3\sigma_{+}) .
\end{equation}  
Here, the transverse effective mass for Ge heavy holes is given by
\begin{equation}
    m_{\text{eff}} = \left(\gamma_1 + \gamma_2 + (2\sqrt{3}\gamma_3)^2\frac{\hbar^2}{2m_e}\sum_{p=2,4,6,\dots} \frac{|\langle 1| k_z|p\rangle ^2}{E_{m} -E_{p,\frac{1}{2}}}  \right) ^{-1}
    \label{eq:meff_ge} ,
\end{equation}
where we define
\begin{gather}
E_m = E_{1,3/2} =  (\gamma_1-2\gamma_2)\dfrac{\hbar^2 \pi^2}{2m_eL_z^2} -\varepsilon , \\
E_{p,3/2} = (\gamma_1-2\gamma_2)\dfrac{\hbar^2 \pi^2p^2}{2m_eL_z^2} -\varepsilon , \\
E_{1,1/2} = (\gamma_1+2\gamma_2)\dfrac{\hbar^2 \pi^2}{2m_eL_z^2} +\varepsilon , \\
E_{p,1/2} = (\gamma_1+2\gamma_2)\dfrac{\hbar^2 \pi^2 p^2}{2m_eL_z^2} +\varepsilon .
\end{gather}
For typical values of $L_z \approx 20$~nm \cite{PhysRevB.103.125201, Wang2024} and $\varepsilon \approx 25$~meV \cite{Wang2024, Wang2021}, we obtain $m_\text{eff}=0.057m_e$, which compares favorably to other estimates of $m_\text{eff}=0.05m_e$ in this system~\cite{PhysRevB.100.041304, PhysRevB.103.125201}.
However, as explained in the main text, we adopt the effective mass for InAs in our simulations.

The Schrieffer-Wolff procedure described above also yields an expression for the Rashba cubic SOC induced by $F_z$:
\begin{align}
    \alpha_C &= -i  \left(\frac{\hbar^2}{2m_e}\right)^2 (eF_z)(\sqrt{3}\bar{\gamma})(2\sqrt{3}\gamma_3)  \notag \\
    &\sum_{p=2,4,6,\dots}2 \langle 1| k_z|p\rangle \langle p| z|1\rangle \left( \frac{\Delta_{m}^{p,\frac{3}{2}}-\Delta_{m}^{p,\frac{1}{2}}+\Delta_{m}^{1,\frac{1}{2}}}{\Delta_{m}^{1,\frac{1}{2}}\Delta_{m}^{p,\frac{1}{2}}\Delta_{m}^{p,\frac{3}{2}}} \right)
    \label{eq:cubic_soc_const_ge} ,
\end{align}
where we define the notation $\Delta_{m}^{i,j} = E_m - E_{i,j} $.
A typical value of electric field $F_z\approx 4.5$~mV/nm \cite{PhysRevB.103.125201} yields a SOC estimate of $\alpha_C \approx 400$~meV$\cdot$nm$^3$, which we adopt throughout this work. 

\section{Origins of spin splitting for a generic dispersion relation} 
\label{AppendixB}

In this Appendix we explore how different dispersion relations, including SOC contributions, like those illustrated in Figs.~\ref{fig:fig3}(b) and \ref{fig:fig3}(c), cause spin splittings to arise between ABSs at the Fermi level.

We first revisit the simple picture of ABSs arising from constructive interference between particle-antiparticle pairs as they traverse a loop in the normal region of the junction, as illustrated in Fig.~\ref{fig:fig3}(a).
This interference condition is expressed in Eq.~(\ref{eq:interference}).
Evaluating this equation near the Fermi level, and making use of Eq.~(\ref{eq:vF}), we obtain
\begin{equation}
 -2\arccos{\left(\frac{E}{\Delta}\right)}\pm\phi + \frac{2lE}{\hbar v_F}= 2\pi n .
 \label{eq:interference2}
\end{equation}
For a given $v_F$, and fixed values of $l$ and $n$, this equation has a single solution, in which $E$ depends implicitly on $v_F$.
However, the dispersion relation has two spin branches, as shown in Figs.~\ref{fig:fig3}(b) and \ref{fig:fig3}(c), which could potentially have different slopes [given by $\hbar v_F=(\partial E/\partial k_x)|_{E=\mu}$], and therefore different energy solutions.

To quantify these solutions, we consider small perturbations to the energy in Eq.~(\ref{eq:interference2}), about the Fermi level: $E=\mu+\delta E$, which arise from small changes in the Fermi velocity $v_F\rightarrow v_{F0}+\delta v$.
A simple expansion in powers of $\delta E$ and $\delta v$ shows how the energy of the ABS is affected by small changes in $v_F$:
\begin{equation}
    \delta E =\left(\frac{2lE}{\hbar v_F^2}\right)
    \left[\frac{2l}{\hbar v_F} + (\Delta^2-E^2)^{-1/2}\right]^{-1}\delta v .
    \label{eq:spin_splitting_deltav_rel}
\end{equation}

We now consider how such changes can arise in a generic 1D dispersion relation of the form
\begin{equation}
    E = Ak_x^2 \pm (Bk_x + Ck_x^3) ,
    \label{eq:csoc_disp}
\end{equation}
where the $A$ term describes the dominant kinetic energy contribution, while the $B$ ($C$) terms describe the much weaker contributions from linear (cubic) SOC.
Below, we ignore branches of the energy dispersion with $k_x<0$ and focus on the two spin branches designated by the $\pm$ sign in Eq.~(\ref{eq:csoc_disp}).

We first consider how SOC affects the dispersion relation near the Fermi level, for a fixed value of $E=\mu$.
Expressing the variations of $k_x$ as $k_x=k_F+\delta k$ and defining $\mu=Ak_F^2$ in the absence of SOC, we obtain the leading-order, SOC-induced corrections,
\begin{equation}
\delta k=\mp (B+Ck_F^2)/2A ,
\label{eq:deltak}
\end{equation}
for the two spin states.
We compute the corresponding distortions of the slope of the dispersion relation, $\hbar v_F=(\partial E/\partial k_x)|_{E=\mu}$, by defining $v_F=2Ak_F+\delta v$.
Collecting linear-order terms and simplifying via Eq.~(\ref{eq:deltak}), we find that
\begin{equation}
\hbar\delta v=\pm 2C\mu/A .
\label{eq:dv}
\end{equation}
In other words, the simple, lateral shift of the dispersion relation caused by linear SOC (the $B$ term) does not affect the slope of the energy dispersion at the Fermi level, and cannot cause a spin splitting; this is an intuitively reasonable result.
However, the cubic SOC (the $C$ term) gives a change of slope, and a corresponding spin splitting, as seen by plugging Eq.~(\ref{eq:dv}) into Eq.~(\ref{eq:spin_splitting_deltav_rel}).

\section{Effective 1D model for cubic and linear SOC}
\label{AppendixA}

In Appendix~\ref{AppendixB}, we showed that the lowest-order (i.e., unrenormalized) linear SOC cannot induce a spin splitting.
As illustrated in Fig.~\ref{fig:fig3}(b), the observed spin splitting is a higher-order effect, requiring couplings between the ground and excited $y$ subbands.
In the current Appendix, we derive this effective coupling for the 1D model described in the main text.
For the case of linear SOC, we show that these $y$ couplings induce an effective cubic SOC term $\alpha_C^*$, which causes a spin splitting through the mechanism described in Appendix~\ref{AppendixB}.

For completeness, we first derive an effective 1D model for the case of cubic SOC.
In Fig.~\ref{fig:fig3}(c), we showed that a spin splitting is already present in this system, at lowest order, through the mechanism described in Appendix~\ref{AppendixB}.
We can therefore obtain an effective 1D model by simply projecting the 2D Hamiltonian model of Eqs.~\eqref{eq:H_tot}-\eqref{eq:H_soc_c}
onto the lowest $y$ subband, yielding
\begin{align}
    H_\text{1D} = \left(\frac{\hbar^2 k_x^2}{2m_\text{eff}} - \mu^* \right)\sigma_0 \tau_z + (\alpha_L^* k_x + \alpha_C k_x^3)\sigma_y\tau_z \nonumber \\
     \quad -\tilde{\Delta}(x)[\tau_{y}\cos{\tilde{\phi}(x)} + \tau_{x}\sin{\tilde{\phi}(x)}]\sigma_{y}
     \label{eq:H1D_cubic_soc2} ,
\end{align}
where $\mu^* = \mu - \frac{\hbar \omega_y}{2}$ and, $\alpha_L^* = \frac{-3\alpha_C m_\text{eff} \omega_y}{2 \hbar}$ are the effective Hamiltonian parameters of this model.
We emphasize again that these parameters do not account for any couplings to excited $y$ subbands.
Consequently, the effective mass and cubic SOC constant are unaffected by the projection procedure.

To make contact, later, with the effective 1D Hamiltonian for the linear SOC system, we perform a further rotation $UH_\text{1D}U^{\dagger}$, using the operator
\begin{equation}
    U = \frac{1}{\sqrt{2}}(\sigma_0\tau_0-i\sigma_x\tau_z) ,
\end{equation}
which represents a combined rotation in particle-antiparticle space and spin space.
The effective 1D Hamiltonian for the case of cubic SOC is then given by
\begin{align}
    H_\text{1D} = \left(\frac{\hbar^2 k_x^2}{2m_\text{eff}} - \mu^* \right)\sigma_0 \tau_z + (\alpha_L^* k_x + \alpha_C k_x^3)\sigma_z\tau_0 \nonumber \\
    \quad -\tilde{\Delta}(x)[\tau_{y}\cos{\tilde{\phi}(x)} + \tau_{x}\sin{\tilde{\phi}(x)}]\sigma_{y} .
    \label{eq:H1D_cubic_soc3}
\end{align}
In Fig.~\ref{fig:fig2}(b), we show that results obtained from this 1D model are in excellent agreement with the full 2D model. 

We now turn to the derivation of an effective 1D model for the case of linear SOC.
Contrary to the cubic SOC case, this requires accounting for couplings between the ground and excited $y$ subbands. 
For clarity, we break the calculation into three steps, as follows. 

\textbf{Step 1: Projecting the 2D Hamiltonian onto the lowest two $y$ subbands.}
As for cubic SOC, we first project the effective 2D Hamiltonian of Eqs.~\eqref{eq:H_tot}-\eqref{eq:H_sc} and \eqref{eq:H_lsoc} onto the $y$ subband basis states defined by Eq.~\eqref{eq:H_0}.
However, unlike cubic SOC, we now include the \emph{two} lowest subband states $\ket{0}$ and $\ket{1}$.
Numerically, we find that limiting our analysis to just these two states is sufficient for capturing the main SOC behavior when $\alpha_L \lesssim 25$ meV$\cdot$nm, as appropriate for the InAs system. 
Within this two-level $y$ subspace, the Hamiltonian may be rewritten as
\begin{equation}
    H_{0} = \left(\frac{\hbar^2k_x^2}{2m_\text{eff}} - \mu  + E_{y+}\right)\sigma_0\lambda_{0}\tau_{z} - (E_{y-}) \sigma_0\lambda_{z}\tau_{z} ,
    \label{eq:H_0_sub} 
\end{equation}
where $\lambda_j$ are Pauli operators acting in the $\ket{0}$ and $\ket{1}$ subband space, $(E_{y+}- E_{y-})$ and $(E_{y+}+E_{y-})$ are the energies of the ground and excited subbands, respectively, and we define $E_{y+} = \hbar\omega_y$ and $E_{y-} = \frac{\hbar\omega_y}{2}$. 
Additionally, the linear SOC can be expressed as 
\begin{equation}
    H_{\text{SOC}} = \alpha_{L}(-k_{x}\sigma_{y}\lambda_{0}\tau_{z} + \kappa\sigma_{x}\lambda_{y}\tau_{0}) ,
    \label{eq:H_lsoc_sub}
\end{equation}
where $\kappa = i\langle0|k_{y}|1\rangle = \sqrt{m_\text{eff}\omega_y/2\hbar}$ , while the superconducting Hamiltonian is given by
\begin{equation}
    H_{\text{SC}} = -\tilde{\Delta}(x)[\cos{\tilde{\phi}(x)}\sigma_{y}\lambda_{0}\tau_{y} + \sin{\tilde{\phi}(x)}\sigma_y\lambda_{0}\tau_{x}] .
    \label{eq:H_sc_sub}
\end{equation}

\textbf{Step 2: Transforming the Hamiltonian into a more convenient form.}
The goal of this Appendix is to perform a perturbation expansion of the Hamiltonian given in Eqs.~(\ref{eq:H_0_sub})-(\ref{eq:H_sc_sub}), to determine the leading-order contribution to the spin splitting.
As explained in Appendix~\ref {AppendixB}, this splitting arises from the cubic SOC constant $\alpha_C^*$ of an effective 1D Hamiltonian, for the case of narrow Josephson junction channels.
Moreover, based on the results reported in Fig.~\ref{fig:fig2}(c), we infer that $\alpha_C^*\propto\alpha_L^5$. 
The inter-subband ($y$) couplings in this system originate from the SOC.
To compute the effective coupling constant $\alpha_C^*$, we could account for these couplings by brute force, for example, by applying a Schrieffer-Wolff transformation. 
However, we already know the answer would require a high-order perturbation analysis ($\alpha_C^*\propto\alpha_L^5$), which is inconvenient.
A more practical approach is to re-express the Hamiltonian in a pseudo-spin basis, formed of a linear combination of the two lowest $y$ subbands, which accounts for a majority of the inter-subband interactions, so that the remaining (very weak) interactions can be treated perturbatively. 
We now identify a set of unitary transformations to obtain the desired pseudo-spin basis.

We first identify the only non-trivial $y$-state coupling terms in Eqs.~(\ref{eq:H_0_sub})-(\ref{eq:H_sc_sub}), which are given by
\begin{equation}
    - (E_{y-}) \sigma_0\lambda_{z}\tau_{z} +  \alpha_L\kappa\sigma_{x}\lambda_{y}\tau_{0} .
    \label{eq:H_ymix} 
\end{equation}
Among these two terms, only the second one couples different $y$ subbands.
As noted above, this mixing term is linear-order in $\alpha_L$ and it is perturbative in the sense that it involves a small dimensionless parameter: $\alpha_L\kappa/E_{y-}\ll 0.1$.
We therefore seek a rotation within the $y$-state subspace that transforms Eq.~(\ref{eq:H_ymix}), such that the prefactor of any resulting $y$-state mixing term ($\sim \lambda_{x,y}$) is proportional to $\alpha_L^2$.

This can be accomplished in two steps.
First, we simplify Eq.~(\ref{eq:H_ymix}) by applying the following rotation to make the mixing term diagonal in spin space:
\begin{equation}
    U = R^{(\sigma)}_y(-\pi/2)\lambda_0\tau_0 .
\end{equation}
This gives
\begin{equation}
    H_{\text{SOC}} = \alpha_{L}(-k_{x}\sigma_{y}\lambda_0\tau_{z} + \kappa\sigma_{z}\lambda_y\tau_{0}) ,
    \label{eq:H_lsoc_t1}
\end{equation} 
while the other terms in the Hamiltonian are unchanged.
Next, we apply a rotation of the form 
\begin{equation}
    U = \cos{\frac{\theta}{2}}\sigma_0\lambda_0\tau_0 +i \sin{\frac{\theta}{2}}\sigma_z\lambda_x\tau_z ,
\end{equation}
where we define  $\tan{\theta} = \alpha_{L}\,\kappa/E_{y-}$.
The transformed Hamiltonian is then given by
\begin{equation}
\begin{aligned}
    H_{0} &= \left(\frac{\hbar^2k_x^2}{2m_\text{eff}} - \mu + E_{y+}\right)\sigma_0\lambda_{0}\tau_{z}  \\
    & \quad - \sqrt{(E_{y-})^2 + (\alpha_{L}\kappa)^2} \sigma_0\lambda_{z}\tau_{z}  , \\
    H_{\text{SOC}} &= -\alpha_{L}k_{x}(\cos{\theta}\sigma_{y}\lambda_{0}\tau_{z} + \sin{\theta}\sigma_x\lambda_{x}\tau_{0}) ,
    \nonumber
    \end{aligned}
\end{equation}
where $H_{\text{SC}}$ is unchanged.
Here, we see that the $y$ coupling term in $H_\text{SOC}$ has the desired prefactor, $\alpha_L\sin\theta\sim\alpha_L^2$.
As a result of these transforms, we only need to apply a low-order perturbation theory to obtain the leading-order spin-splitting results.

\textbf{Step 3: Schrieffer-Wolff transformation.}
We finally apply a Schrieffer-Wolff transformation to the previous Hamiltonian, up to second order in the inter-subband coupling constant, to block-diagonalize the two-subband system.
We then truncate the lowest $y$ subband, yielding our final result for the effective 1D Hamiltonian, for the case of linear SOC:
\begin{align}
    H_\text{1D} = \left(\frac{\hbar^2k_x^2}{2m^*_\text{eff}} - \mu^*\right)\sigma_0 \tau_z + (\alpha_L^* k_x + \alpha_C^* k_x^3)\sigma_z\tau_0 \nonumber \\
    \quad -\tilde{\Delta}(x)[\tau_{y}\cos{\tilde{\phi}(x)} + \tau_{x}\sin{\tilde{\phi}(x)}]\sigma_{y} .
\end{align}
Here, the renormalized Hamiltonian parameters are all computed at the same level of approximation, given by
\begin{equation}
\begin{aligned}
    \frac{\hbar^2}{2m_\text{eff}^*} &= \frac{\hbar^2}{2m_\text{eff}} - \frac{\alpha_L^4\kappa^2}{2E_{y-}^3} + \mathcal{O}(\alpha_L^6) , \\
    \mu^* &= \mu - \frac{\hbar\omega_y}{2} + \frac{\alpha_L^2\kappa^2}{2E_{y-}}-\frac{\alpha_L^4\kappa^4}{8E_{y-}^3} + \mathcal{O}(\alpha_L^6), \\
    \alpha_L^* &= -\alpha_L + \frac{\alpha_L^3\kappa^2}{2E_{y-}^2}-\frac{3\alpha_L^5\kappa^4}{8 E_{y-}^4} + \mathcal{O}(\alpha_L^6),\\
    \alpha_C^* &= \frac{\alpha_L^5\kappa^2}{2E_{y-}^4} + \mathcal{O}(\alpha_L^6).
\end{aligned}
\label{eq:fifthpower}
\end{equation}
Hence, we confirm that $\alpha_C^* \propto \alpha_L^5 $, within the approximations considered here.

\begin{figure*}[t]  
    \centering
    \includegraphics[width=0.75\textwidth]{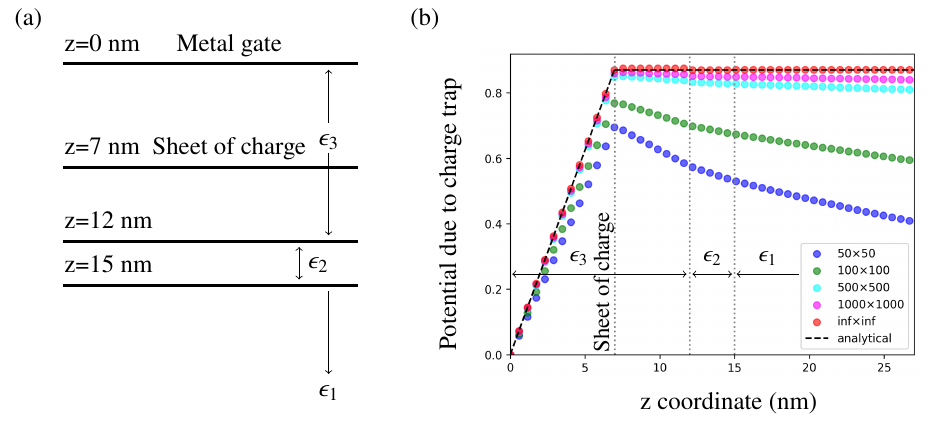} 
     \caption{
     Checking the accuracy of the image charge method used to compute the electrostatic potential from a trapped charge.
     (a) Schematic cross-section of the structure considered in this analysis, with interface positions $z$.
     The heterostructure is comprised of three layers, identified by their different dielectric constants, capped by a uniform metal top gate.
     Here, we assume $\epsilon_3 = 8, \epsilon_2=12.75$, and $\epsilon_1=16.2$, corresponding to the materials Al$_2$O$_3$, SiGe, and Ge. 
     A uniform sheet of charge of areal density $\sigma$ is positioned inside the top layer.
     (b) Solutions for the electrostatic potential as a function of the vertical position $z$, obtained for the geometry shown in (a), assuming a grounded top gate.
     Results are reported in units of $\sigma/\epsilon_0$.
     The dashed line describes the exact analytical solution.
     Colored markers show solutions obtained using the Green's function method described in Appendix~\ref{AppendixC}, obtained for different integration ranges [see Eq.~(\ref{Vqw})], as indicated in the legend.
     The red markers correspond to an effectively infinite domain, as computed using the \texttt{scipy.integrate} routine.
     The convergence of the numerical and analytical results for large integration domains demonstrates the accuracy of the image charge method.
    }
    \label{fig:fig5}
\end{figure*}

\section{Modeling charge traps} 
\label{AppendixC}
In Figs.~\ref{fig:fig4}(a)-\ref{fig:fig4}(c), we study the effects of charge traps on the spin splitting of ABSs, in planar Josephson junctions, with linear and cubic SOC.
In other sections of this work, we have described the quantum mechanical models used to solve the spin splitting. 
In this Appendix, we describe the electrostatic model of a charge trap, which we then incorporate into the quantum calculations.

We consider the heterostructure whose cross-section is shown in Fig.~\ref{fig:fig4}(a) (not to scale).
As explained in the main text, the charge trap is assumed to reside at the SiGe-oxide interface.
For definiteness, we adopt heterostructure parameters consistent with recent spin-qubit experiments in a Ge quantum well \cite{Hendrickx2024}.
The relevant materials layers include 7~nm of oxide, a 48~nm  spacer layer of SiGe, a 20~nm Ge quantum well, and a semi-infinite layer of SiGe.
As illustrated in Fig.~\ref{fig:fig4}(a), the quantum well is sandwiched between SiGe on either side.
The relative dielectric constants in the three layers are given by 8, 12.75 and 16.2, respectively \cite{Schaffler1997}.
The top gate in this structure is taken to be an infinite metal sheet, with Dirichlet boundary conditions (i.e., with perfect screening).
We then compute the electrostatic potential at the center of mass of the 2DHG wave function, which is taken to lie 2~nm below the upper Ge/SiGe interface.
To provide an apples-to-apples comparison between the linear and cubic SOC simulations, we use the same electrostatic potential in both cases. 

A common method for solving problems of this type is the method of images.
A simple approach is to treat the various layers as if they were a uniform material, with a single dielectric constant and a metal top gate \cite{PhysRevLett.128.247701,Wang2024}.
The resulting solution involves only a single image charge.
However, this approximation may be inaccurate, especially for a large dielectric mismatch between the oxide and semiconducting layers.
A more accurate treatment should include an infinite series of image charges, with magnitudes chosen to satisfy the boundary conditions at each internal interface.
For the heterostructure we consider, which includes a metal top gate and three internal interfaces, the full set of images is somewhat complicated.
We therefore adopt the rather efficient method of images described in Ref.~\cite{ct_comp}.
As in other schemes, the image charges are determined recursively, up to a desired level of accuracy.
The final electrostatic potential is then expressed as a sum of contributions from an appropriate set of images.
Although the method is described in detail in Ref.~\cite{ct_comp}, we summarize it here, for completeness.
For convenience, we also adopt the notation of Ref.~\cite{ct_comp}, while interchanging $x\leftrightarrow z$.

We define the oxide layer as medium III, the SiGe layer as medium II, and the (semi-infinite) Ge quantum well as medium I, with corresponding dielectric constants $\epsilon_3=8, \epsilon_2=12.75$, and $\epsilon_1=16.2$.
The oxide, SiGe and Ge layers have thicknesses $c=7$~nm, $b=48$~nm and $a=20$~nm respectively. 
The last SiGe layer is treated as semi-infinite. 
The charge trap is assumed to reside in medium III, very near or directly on the SiGe/oxide interface, while the electrostatic potential is eventually evaluated in medium I.
In the scheme of \cite{ct_comp}, the effect of the images can be viewed as analogous to transmitted and reflected light rays at the internal boundaries.
The image charges needed to satisfy these transmission and reflection conditions are classified as follows. 
The images responsible for the total reflection at the metal interface in medium III are labeled $q_{4j}$.
The images causing reflection into medium III from the SiGe/oxide interface are labeled $q_{3j}$.
The images responsible for transmission into medium II at this same interface are labeled $q_{2j}$.
The images causing reflection into medium II from the SiGe/Ge interface are labeled $q_{1j}$.
The images responsible for transmission into medium I at this interface are labeled $q_{5j}$.

The recursion proceeds as follows.
An initial source charge $Q$ is chosen, with a location given by $(x,y,z)=(0,y_\text{ct},z_\text{ct})$.
In our problem, we take $Q=|e|$ for the case of Ge and $Q=-|e|$ for the case of InAs.
The effect of this initial source charge is to induce a reflection or transmission at a given interface.
The image charge associated with this reflection or transmission is computed as a function of the original source charge.
The resulting image charges are then taken as new source charges, and the procedure is repeated until the magnitude of the image charge is less than a specified fraction of the original charge $Q$. 
For our calculations, we choose this convergence condition to be $|q_{\gamma j}|<10^{-5}|Q|$.
In the notation of \cite{ct_comp}, the index $j$ of a given image charge $q_{\gamma j}$ simply reflects the number of reflections/transmissions of type $\gamma$ that occur before the image $q_{\gamma j}$ is generated.

Taking an example, the initial conditions for our recursion procedure are given by $q_{30}=Q$ and $z_{30}=z_\text{ct}$.
(Here, we define the oxide-metal interface location to be at $z=0$.
For a charge trap at the SiGe/oxide interface, we then have $z_\text{ct}=c$.)
This source charge causes a pair of reflections back into medium III via the image charges $q_{41}$ and $q_{31}$ at locations $z_{41}$ and $z_{31}$, a transmission into medium II via the image charge $q_{21}$ at location $z_{21}$, with a further reflection and transmission at the SiGe/Ge interface via the image charges $q_{11}$ and $q_{51}$ at locations  $z_{11}$ and $z_{51}$, respectively.
The full set of recursion relations is given as follows, with the sources listed on the right-hand side of the equations, and the corresponding images listed on the left-hand side:
\begin{gather}
    q_{2j} = \alpha_{23}q_{1i}, \quad z_{2j} = -z_{1i} + 2c , \\
    q_{3k} = \beta_{23}q_{1i}, \quad z_{3k} = z_{1i}  , \\
    q_{1j} = \alpha_{21}q_{2i}, \quad z_{1j} = -z_{2i} + 2(b+c) , \\
    q_{5k} = \beta_{21}q_{2i}, \quad z_{5k} = z_{2i}  , \\
    q_{4j} = -q_{3i}, \quad z_{4j} = -z_{3i} , \\
    q_{3j} = \alpha_{32}q_{4i}, \quad z_{3j} = -z_{4i} + 2c , \\
    q_{2k} = \beta_{32}q_{4i}, \quad z_{2k} = z_{4i} , \\
    q_{6j} = \alpha_{10}q_{5i}, \quad z_{6j} = -z_{5i} + 2(a+b+c) , \\
    q_{7k} = \beta_{10}q_{5i}, \quad z_{7k} = z_{5i} , \\
    q_{5j} = \alpha_{12}q_{6i}, \quad z_{5j} = -z_{6i} + 2(b+c) , \\
    q_{1k} = \beta_{12}q_{6i}, \quad z_{1k} = z_{6i} . 
    \label{eq:ic2}
\end{gather}
Here, the indices $j$ and $k$ follow the sequential ordering scheme mentioned above, for images of a given type.
We have also defined
\begin{multline}
    \begin{aligned}
        \alpha_{ij} = \frac{\epsilon_i - \epsilon_j}{\epsilon_i+\epsilon_j} , \quad \beta_{ij} = 1-\alpha_{ij} .
    \end{aligned}
    \label{eq:ic5}
\end{multline}

Finally, the electrostatic potential for a charge trap in the Ge layer is given by
\begin{multline}
    V_\text{ct}(x,y,z) \\
    = \sum_i \frac{q_{5i}}{4\pi\epsilon_0\epsilon_1\sqrt{x^2 + (y-y_\text{ct})^2 + (z-z_{5i})^2}} .
\label{Vqw}
\end{multline}
We then evaluate $V_\text{ct}$ at the center of mass of the 2DHG, given by $z=b+c+2$~nm.
Additional details can be found in Ref.~\cite{ct_comp}.

To verify this image charge method, we now use it to solve a problem whose solution can be computed analytically, namely, the case of a heterostructure containing an infinite (uniform) sheet of charge.
In this case, the geometry is effectively 1D, which allows us to obtain a simple analytical solution via elementary techniques.
The image charge solution for this geometry is obtained by modifying Eq.~(\ref{Vqw}) to describe a test charge of unit magnitude ($Q=1$), at location $(x,y,z)=(x',y',z'=7~\text{nm})$, in the same heterostructure.
This solution represents a Dirichlet Green's function, defined as $G_D(x,y,z,x',y',z')$.
We can then solve for the electrostatic potential arising from a sheet of charge, with areal density $\sigma$, by performing the integral
\begin{equation}
    \phi(x,y,z)  = \sigma\int G_{D}(x, y, z, x', y', z') dx' dy' .
    \label{eq:GreensFn}
\end{equation}

We consider the specific geometry shown in Fig.~\ref{fig:fig5}(a), comprising a top gate, three dielectric layers, and a charge sheet.
The analytical solution for this problem is solved using a conventional Gaussian pill-box method, taking as a boundary condition that the metal layer, located at $z=0$, is grounded.
The resulting potential is plotted as a dashed line in Fig.~\ref{fig:fig5}(b), as a function of vertical position.
We also solve the integral in Eq.~(\ref{eq:GreensFn}) numerically, noting that (for this problem) the result is quite sensitive to the integration range. 
In Fig.~\ref{fig:fig5}(b), we plot the results of this numerical calculation, with different colored markers representing different integration ranges.
We find that the numerical solution converges nicely to the analytical solution, for large integration ranges, demonstrating the accuracy of the image charge technique.

\bibliography{references}          

\end{document}